\documentclass[twocolumn,showpacs,prd,aps,superscriptaddress, amsmath, preprintnumbers,amssymb]{revtex4-2}
\usepackage{graphicx} 
\usepackage{todonotes}
\usepackage{amsmath}
\usepackage{multirow}
\usepackage{makecell}
\usepackage{amssymb}
\usepackage{hyperref}
\usepackage{physics}

\usepackage{fontawesome}

\usepackage{tikz}
\usetikzlibrary{shapes}
\usetikzlibrary{positioning,
                shadows,
                shapes.geometric}
\usepackage{xcolor}
\definecolor{eorange}{RGB}{230, 159, 0}
\definecolor{eblue}{RGB}{86, 180, 233}

\usepackage{xspace}
\usepackage{orcidlink}

\usepackage{soul}

\newcommand{\newAdaptName}{(SC)$^2$-ADAPT-VQE\xspace}
\newcommand{\grdapx}{|\psi_{\rm sur}\rangle}
\newcommand{\Adapt}{ADAPT-VQE\xspace}

\newcommand{\SCAdapt}{SC-ADAPT-VQE\xspace}

\newcommand{\chiconexp}{\langle \bar{\psi}\psi\rangle}
\newcommand{\new}[1]{\textcolor{blue}{#1}}

\newcommand{\BNL}{Department of Physics, Brookhaven National Laboratory, Upton New York}
\newcommand{\USRA}{Universities Space Research Association, Research Institute for Advanced Computer Science (RIACS) at NASA Ames Research Center, Mountain View California}
\newcommand{\NASA}{NASA Ames Research Center, Moffet Field California}

\newcommand{\VTQ}{Virginia Tech Center for Quantum Information Science and Engineering, Blacksburg, VA 24061}
\newcommand{\VTPhysics}{Department of Physics, Virginia Tech, Blacksburg, VA 24061}
\newcommand{\VTChemistry}{Department of Chemistry, Virginia Tech, Blacksburg, VA 24061}
\newcommand{\FSU}{Department of Physics, Florida State University, Tallahassee, FL 32306}
\newcommand{\IA}{Iowa State University, Ames Iowa}
\newcommand{\FNAL}{Fermi National Accelerator Laboratory, Batavia,  Illinois, 60510, USA}
\newcommand{\SQMS}{Superconducting and Quantum Materials System Center (SQMS), Batavia, Illinois, 60510, USA}

  \colorlet{AFc}{purple!50!white}

  \newcommand{\Eq}[1]{Eq.\,\eqref{eq:#1}}
  \newcommand{\Eqs}[1]{Eqs.\,\eqref{eq:#1}}
  \newcommand{\Fig}[1]{Fig.\,\ref{fig:#1}}

\begin{document}

\preprint{FERMILAB-PUB-24-0456-SQMS-T}

\begin{abstract}
Inspired by recent advancements of simulating periodic systems on quantum computers,  we develop a new approach, \newAdaptName,
to further advance the simulation of these systems. 
Our approach extends the scalable circuits \Adapt framework, which builds an ansatz from a pool of coordinate-invariant operators defined for arbitrarily large, though not arbitrarily small, volumes. Our method uses a classically tractable ``Surrogate Constructed'' method to remove irrelevant operators from the pool, reducing the minimum size for which the scalable circuits are defined.  Bringing together the scalable circuits and the surrogate constructed approaches forms the core of the (SC)$^2$ methodology.
    Our approach allows for a wider set of classical computations, on small volumes, which can be used for a more robust extrapolation protocol. While developed in the context of lattice models, the surrogate construction portion is applicable to a wide variety of problems where information about the relative importance of operators in the pool is available. 
        As an example, we use it to compute properties of the  
    Schwinger model --- quantum electrodynamics for a single, massive fermion in $1+1$ 
    dimensions --- and show that our method can be used to accurately extrapolate to the continuum limit. 
\end{abstract}

\title{Surrogate Constructed Scalable
Circuits
ADAPT-VQE in the Schwinger model}

\author{Erik Gustafson${}^{\orcidlink{0000-0001-7217-5692}}$}
\thanks{Erik Gustafson and Kyle Sherbert contributed equally.}
\affiliation{\USRA}

\author{Kyle Sherbert${}^{\orcidlink{0000-0002-5258-6539}}$}
\thanks{Erik Gustafson and Kyle Sherbert contributed equally.} 
\affiliation{\VTChemistry} \affiliation{\VTPhysics} \affiliation{\VTQ}

\author{Adrien Florio${}^{\orcidlink{0000-0002-7276-4515}}$}
\affiliation{\BNL}

\author{Karunya Shirali${}^{\orcidlink{0000-0002-2006-2343}}$}
\affiliation{\VTPhysics} \affiliation{\VTQ}

\author{Yanzhu Chen$^{\orcidlink{0000-0001-5589-9197}}$}
\affiliation{\VTPhysics} \affiliation{\VTQ}
\affiliation{\FSU}

\author{Henry Lamm${}^{\orcidlink{0000-0003-3033-0791}}$}
\affiliation{\SQMS}
\affiliation{\FNAL}

\author{Semeon Valgushev$^{\orcidlink{https://orcid.org/0000-0002-4306-1423}}$}
\affiliation{\IA}

\author{Andreas Weichselbaum$^{\orcidlink{0000-0002-5832-3908}}\,$}
\affiliation{\BNL}

\author{Sophia E. Economou${}^{\orcidlink{0000-0002-1939-5589}}$}
\affiliation{\VTPhysics} \affiliation{\VTQ}

\author{Robert D. Pisarski${}^{\orcidlink{0000-0002-7862-4759}}$}
\affiliation{\BNL}

\author{Norm M. Tubman${}^{\orcidlink{0000-0002-9577-8485}}$}
\affiliation{\NASA}
\email{norman.m.tubman@nasa.gov}

\date{\today}

\maketitle

\section{Introduction}

State preparation is a crucial component of many quantum algorithms, including phase estimation and time evolution, in all areas of physical simulation from chemistry, to materials science, to high energy and nuclear physics \cite{bauer_quantum_2020,mcardle2020quantum,cao2019quantum,Grabowska:2024vop,Klco:2021lap,Alam:2022crs,Feynman:1981tf, Bauer:2022hpo, DiMeglio:2023nsa,Bauer:2023qgm,PRXQuantum.3.020323,stateprep14}.
Significant work has been done on state preparation for materials and chemical systems~\cite{Aspuru-Guzik2005, peruzzo2014variational, Bauer2016, Berry2018, Yuan2019,stateprep11, motta2020determining, Gomes2021, Larsen2023,stateprep12,stateprep13},
and more recently in nuclear and high energy physics \cite{Kokail:2018eiw,Gustafson:2020yfe,Turro:2021vbk,Ciavarella:2021lel,Farrell:2023fgd,Belyansky:2023rgh,Lee:2023urk,Farrell:2024fit,Li:2024lrl,Turro:2024pxu}.
Notably, the Variational Quantum Eigensolver (VQE) can prepare ground states using shallow depth circuits~\cite{peruzzo2014variational,McClean2016,cerezo2021variational,tilly_variational_2022,fedorov2022vqe,stateprep15}, and
in particular, the Adaptive Derivative-Assembled Problem Tailored VQE (\Adapt) \cite{Grimsley:2018wnd,Grimsley_2023} shows great promise in finding ground states with compact circuits for a wide variety of problems.
In this algorithm, the variational quantum circuit is iteratively built by exponentiating simple operators greedily selected from a user-defined pool, where the coefficient for each operator is treated as a variational parameter.
The operators in the pool determine the expressivity of the ansatz, while the size of the pool determines the measurement overhead required for implementation.

Systems in materials science and nuclear physics often have discrete translational invariance (i.e. periodicity).
These systems are typically very challenging, since very large volumes are required to obtain physically meaningful results.
However, variational algorithms can incorporate translational invariance with a variety of strategies, compatible with \Adapt.
For example, one could adopt a momentum basis, such that all operators implicitly satisfy translational invariance~\cite{manrique2021}. 
Alternatively, one could use a local real-space basis, and ``tile'' operators to build up an operator pool,
such that local interactions relevant in small volumes are systematically integrated into larger volumes~\cite{OperatorTiling2024}.
Finally, one can enforce translational invariance in the real-space representation.
This is the approach taken by \cite{Farrell:2023fgd, Farrell:2024fit}, which introduced the ``Scalable Circuits'' \Adapt variant, or \SCAdapt.

In \SCAdapt, one first solves the \Adapt problem for small volumes which are classically simulable, to obtain both a sequence of (translationally-invariant) operators and the optimal coefficients for each operator.
One then extrapolates the coefficients to the thermodynamic (large-volume) limit, to obtain a circuit that can be run, without any further optimization, on quantum hardware.
Refs. \cite{Farrell:2023fgd, Farrell:2024fit} applied \SCAdapt to study the Schwinger model \cite{Schwinger:1962tp, Coleman:1975pw, Coleman:1976uz}, a prototypical theory of gauge fields with matter, to demonstrate a procedure of scalable circuits.
Their work considered systems of sizes up to 32 qubits as ``small'', and demonstrated circuit viability on QPUs for ``large'' systems with over 100 qubits.

While the results in Refs. \cite{Farrell:2023fgd,Farrell:2024fit} are impressive,
\Adapt will tend to select a different sequence of operators for each different volume.  This is
especially true for the later operators needed to obtain high-accuracy trial states, and this can result in ambiguities 
extrapolating to large volumes.
Furthermore, it is not yet clear whether these methods can be used to explore the regime where the lattice spacing goes to zero, an essential step to ensure that a lattice theory can be used to study the corresponding field theory in the continuum limit.
Therefore, two crucial questions remain:
\begin{enumerate}
    \item How can the \SCAdapt framework be modified to obtain less ambiguous extrapolations to the thermodynamic limit?
    \item Can the \SCAdapt framework be used to obtain accurate extrapolations to the continuum limit?
\end{enumerate}

To address the first question, we introduce  Surrogate-Constructed Scalable-Circuits \Adapt, or \newAdaptName.
This new variant modifies \SCAdapt in two ways:
First, while one adopts the same translationally-invariant operator pool as in Refs. \cite{Farrell:2023fgd,Farrell:2024fit}, one uses a large-volume classical surrogate such as matrix product states (used in this work), density matrix renormalization groups, or classical coupled cluster equations 
to pre-screen the most relevant operators, reducing the size of the pool.
Second, one solves \Adapt~ {\em just once}, on the largest classically emulateable volume, and then optimizes the coefficients for the {\em same operator sequence} on each smaller volume.

To address the second question, we use our \newAdaptName method to perform extrapolations to the infinite volume (thermodynamic) limit, for a series of systems with reduced lattice spacing.
We then extrapolate the thermodynamic limit results for each problem instance to the continuum limit, where the lattice spacing vanishes.
We demonstrate that, even in the absence of large-volume runs on quantum hardware, the \newAdaptName ansatz provides a robust method to calculate observables in the continuum limit.

This paper is organized as follows. 
In Sec. \ref{sec:latticeschwingermodel} we discuss the target Hamiltonian for our studies, which is that of the Schwinger model, Refs. \cite{Farrell:2023fgd,Farrell:2024fit}.
We elaborate on VQE, \Adapt, and \SCAdapt, in Sec. \ref{sec:adaptvqe}.
We describe our own method, \newAdaptName, in Sec.~\ref{sec:sc2}.
The results from our simulations, including extrapolations to the continuum limit for the massless Schwinger model, are provided in Sec. \ref{sec:results}.
In Sec. \ref{sec:outlook} we discuss the impact of this work, and future directions.

\section{Lattice Field Theories and the Schwinger Model}
\label{sec:latticeschwingermodel}

Before discussing our particular system, we offer a brief 
review of quantum computing for quantum field theories.
The principal method for quantum simulations of field theories utilizes lattice models. In this method, the infinite dimensional Hilbert space is partially truncated by only utilizing a finite number of points in space\new{,} while leaving the remaining degrees of freedom unchanged; this leaves a finite number of locally infinite dimensional Hilbert spaces. A key requirement for quantum simulation is mapping the infinite dimensional Hilbert space to finite dimensional quantum resources.
Significant work has been done to progress simulations of these models across a wide range of digitizations such as discrete groups \cite{Gustafson:2022xdt, Gustafson:2023kvd,Alam:2021uuq,Ji:2020kjk,Armon:2021uqr,Ji:2022qvr,Alexandru:2019nsa,Alexandru:2021jpm,Charles:2023zbl,Murairi:2024xpc,Gustafson:2024kym,Assi:2024pdn,Lamm:2024jnl}, loop-string-hadron formulations \cite{Raychowdhury:2018osk,Raychowdhury:2019iki,Davoudi:2020yln,Kadam:2023gli,Davoudi:2020yln,Mathew:2022nep,Kadam:2024zkj}, quantum link models \cite{Liu:2021tef,Brower:1997ha,Brower:2020huh,Chandrasekharan:1996ih,Zhou:2021qpm}, electric field truncations \cite{Klco:2018kyo,Klco:2019evd,Klco:2019evd, Farrell:2023fgd, Farrell:2024fit, Illa:2024kmf,Ciavarella:2021nmj, Bazavov:2015kka, Zhang:2018ufj,PhysRevD.99.114507,Bauer:2021gek,Grabowska:2022uos,Buser:2020uzs,Bhattacharya:2020gpm,Kavaki:2024ijd,Calajo:2024qrc,Murairi:2022zdg,Zohar:2015hwa,Zohar:2012xf,Zohar:2012ay,Zohar:2013zla} and other novel formulations
\cite{Barata:2020jtq,Zohar:2014qma,Kreshchuk:2020dla,Kreshchuk:2020aiq,Kreshchuk:2020kcz,Liu:2020eoa,Fromm:2023bit, Li:2024ide,Hardy:2024ric}. 
Additional work has been done on simulation methods for measuring observables \cite{Barata:2024bzk,Gustafson:2019mpk,Cohen:2021imf,Lamm:2019uyc,Gustafson:2023ayr,Turro:2024pxu}, road maps and resource estimates for simulations \cite{Ciavarella:2021nmj,Zohar:2012ay,Zohar:2021nyc,Bender:2018rdp,Farrell:2022wyt,Farrell:2022vyh}, proposals for simulations on hardware \cite{Lu:2018pjk,kurkcuoglu2022quantum,Zohar:2015hwa,Zohar:2013zla,Zohar:2016iic,Zohar:2012xf}, multigrid methods for quantum simulation\cite{Illa:2022jqb}, and state preparation techniques \cite{Popov:2023xft,Farrell:2023fgd,Klco:2019xro,Halimeh:2020kyu,Farrell:2024fit,Farrell:2024mgu,Gustafson:2020yfe,Gustafson:2020vqg}. 
Finally, there has also been complementary development of simulation tools and uses for quantum computation of field theories such as classical and classical-quantum feedback \cite{Avkhadiev:2019niu, Carena:2021ltu,Carena:2022hpz, Zohar:2018nvl,Avkhadiev:2022ttx,Fleming:2024kss,Crippa:2024cqr,Guo:2024tnb}, error mitigation and correction schemes \cite{1797835,Carena:2024dzu,Halimeh:2020xfd,Halimeh:2021vzf,Tran:2020azk,Halimeh:2019svu,Halimeh:2020djb, Halimeh:2020ecg,Kasper:2020owz}, more accurate Hamiltonians \cite{Carena:2022kpg,Ciavarella:2024fzw, Gustafson:2022hjf}, improved circuits \cite{ffttocome}, reduced memory requirements \cite{Zohar:2019ygc, Zohar:2018cwb},
uses for qudits 
\cite{Gustafson:2021qbt,Gustafson:2022xlj,illa2024qu8its}, 
and other techniques and benchmarks \cite{Gustafson:2021jtq,Gustafson:2019vsd}.

We use the lattice Schwinger model as a prototype field theory to showcase our \newAdaptName algorithm, just as it was used by Refs. \cite{Farrell:2023fgd, Farrell:2024fit} to showcase \SCAdapt.
The lattice Schwinger model is a spatial discretization of quantum electrodynamics in $1+1$ dimensions, with a single species of fermion. 
Due to its relative simplicity,
this model from high energy physics
has been studied extensively as a test case for quantum simulations \cite{PhysRevD.105.094503,Davoudi:2022uzo,Honda:2021aum,Mueller:2022xbg,davoudi2024scattering,PhysRevD.106.054508,PhysRevD.106.054509,Shaw:2020udc,Farrell:2024mgu,Florio:2024aix,Papaefstathiou:2024zsu}. 
In fact it is closely related to a type of Heisenberg spin chain with alternating external magnetic fields.
After integrating out the bosonic degrees of freedom and performing a Jordan-Wigner transformation, the Hamiltonian is
\begin{align}
\label{eq:schwingerham}
    \hat{H} = &\frac{1}{2a} \sum_{j=0}^{N_s - 2} (\hat{\sigma}^+_j \hat{\sigma}^-_{j+1} + \hat{\sigma}^+_{j+1}\hat{\sigma}^-_{j})+ \frac{m_0}{2} \sum_{j=0}^{N_s-1}(-1)^j\hat{\sigma}^z_j\notag\\
    & + \frac{a g^2}{8} \sum_{j=0}^{N_s - 2} (\sum_{k=0}^{j}(\hat{\sigma}^z_k + (-1)^k))^2,
\end{align}
where $m_0$ is the bare mass of the fermion, $g$ is the 
bare electric charge (coupling constant) for the gauge field
\footnote{
For faster convergence to the continuum limit,
one may replace $m_0 \to m_{lat.} \equiv m_0 - \frac{a g^2}{8}$
in \Eq{schwingerham} 
which
greatly reduces errors from discretization
 \cite{Dempsey:2022nys}.
 In this work we ignore this correction, 
 simply to provide a demonstration of the efficacy of \newAdaptName.},
 and $N_s$ is the number of sites.  With Kogut-Susskind (frequently referred to as staggered) fermions,
 $N_s$ must be an even integer. 

Note that in the rest of this work, we will consider the {\em massless} Schwinger model, in which $m_0=0$. The continuum model is exactly solvable and equivalent to a $massive$ free bosonic model \cite{Coleman:1976uz}. This makes it particularly appealing as a test case, as the whole spectrum of the theory is known analytically. Notably, the first excited state has a mass of $m_{\rm gap}=g/\sqrt{\pi}$ or conversely a correlation length of $\xi\sim \sqrt{\pi}/g$. This gives an estimate of how much correlation needs to be built-into the system and how large a system is needed to correctly capture the infinite volume system.
The correlation length gives an approximate estimate of how far interactions typically extend. If the physical volume, here the length $N_s$, is too small compared to correlation length, then there will be noticeable self interaction effects or boundary contamination. Thus a sufficiently large volume is one that is multiple times the correlation length of the system. 

\section{Existing Methods}
\label{sec:adaptvqe}

\begin{figure*}
    \includegraphics[width=\linewidth]{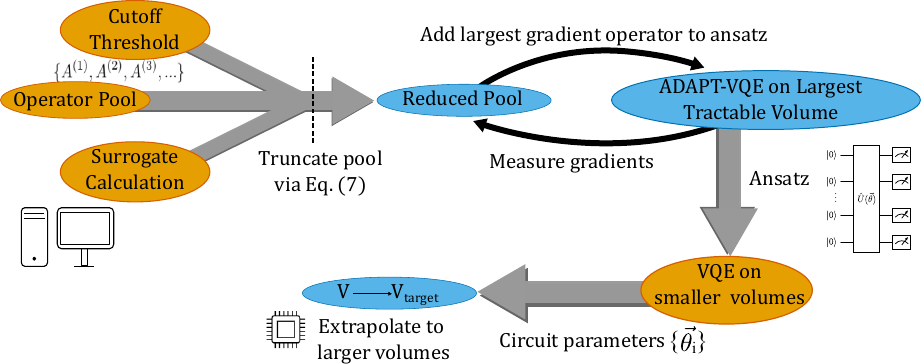}
    \caption{Flow chart describing the \newAdaptName work flow.
     Blue components indicate the workflow with
     (SC-)ADAPT-VQE 
     \cite{Farrell:2023fgd,Farrell:2024fit} while orange
     components those added with \newAdaptName.
     The surrogate calculations along with a cutoff threshold, $\Delta$ come together to give a rigorous definition of a truncated operator pool. The reduced pool is utilized in an \Adapt calculation at the largest volume that can be simulated on a classical computer. This ansatz is then optimized on smaller volumes.
     If so desired, the resulting parameters $\{\vec{\theta}_i\}$
     can then be extrapolated to larger
     volumes on a quantum computer. The thin black
     arrows
     illustrate that the classical adaptive part of ADAPT-VQE iteratively constructs
     a quantum circuit, using operators sampled from the truncated pool.
     }
    \label{fig:daflowchart}
\end{figure*}

The VQE \cite{peruzzo2014variational} is a state preparation algorithm that begins with some initial state $|\psi_{\rm ref}\rangle$ which is trivial to initialize on a quantum computer,
and then constructs an optimal trial state
$|\psi\rangle = \hat U(\vec{\theta}) |\psi_{\rm ref}\rangle$,
where $\hat U$ is a unitary operator encoded by a quantum circuit parameterized by angles $\vec{\theta}$.
A classical optimization is used to find the ideal choice of angles that minimize some objective function.
For example, to prepare the ground state of a physical system
the objective function can be the expectation value of the energy.
In this paper we assume the latter whenever we refer
to optimization. To be specific, our objective function
is $E_\psi \equiv \langle\psi| H|\psi \rangle$,
using the Hamiltonian of \Eq{schwingerham}.
We note that, especially for one-dimensional systems with finite entanglement entropy, an alternative objective function for designing scalable quantum circuits could be the overlap with the ground state of a classical surrogate, such as an MPS \cite{Feniou_2023}. In this case one would use the fidelity of the target state, which might speed up the preparation of excited quantum states.

\Adapt~\cite{Grimsley:2018wnd} is an 
algorithm that iteratively extends and updates
the unitary
\begin{equation}
    \label{eq:vqeansatz}
    \hat U^{(k)}(\vec{\theta}^{(k)}) = \prod_{j=1}^k e^{-i \theta_j^{(k)} \hat{O}_{i_j}}, 
\end{equation}
where each element $\theta_j^{(k)}$ of the parameter vector $\vec{\theta}^{(k)}$ is the coefficient corresponding to the operator $\hat{O}_{i_j}$ drawn from a predefined pool $\{\hat O_i\}$.
Note that $\hat U^{(k)}$ encodes the entire circuit
up to iteration $k$, where
each $\theta_j$ is equivalent to a time over which the trial state is evolved under a Hamiltonian
$\hat{O}_{i_j}$.
In each iteration of \Adapt, the unitary $\hat U$ is extended with a new $\hat{O}_{i_{k+1}}\in\{\hat O_i\}$ using a ``greedy'' procedure,
selecting that $\hat{O_i}$ which confers the largest gradient $|\partial E_\psi/\partial \theta_{k+1}|_{\vec \theta^{(k)}}$ of the objective function at the current optimized trial state.
After appending the new operator, the variational parameters $\theta_j^{(k+1)}$ in the circuit are reoptimized, initializing the values $\theta_{j\le k}$ with the optimized value from the previous iteration, and initializing $\theta_{k+1}$ to zero.
Operators are appended and optimizations repeatedly, until the gradients corresponding to every candidate
operator
fall below some user-defined
threshold $\epsilon$,  such that
\begin{equation}
\label{eq:adapteps}
  \epsilon >
  \left|\frac{\partial E_\psi}{\partial \theta_{k+1}}\right|_{\vec \theta^{(k)}_{\rm converged}} \; ,
\end{equation}
for all choices $\hat O_{i_{k+1}} \in \{\hat O_i\}$. 
Note that the \Adapt convergence criterion, $\epsilon$, is independent of, and typically larger than, the convergence criterion used to terminate each individual optimization.

For small molecules mapped to fewer than 14 qubits, this has been shown to yield highly compact ans\"atze and reach arbitrarily accurate energies~\cite{Grimsley:2018wnd}.

While the original \Adapt algorithm used fermionic single- and double-excitations as the pool operators, there has been significant work to find optimal pool operator choices for \Adapt.
For instance, Ref.~\cite{Tang2021} introduces the qubit-ADAPT operator pool, which breaks Jordan-Wigner-transformed fermionic excitations into individual terms, and drops $Z$-operators to make them more hardware-efficient.
Ref.~\cite{Yordanov2021} introduces qubit-excitation based operators, another hardware-efficient construction in which $Z$-operator strings are removed from the Jordan-Wigner-transformed fermionic excitations.
Ref.~\cite{Shkolnikov2021} discusses symmetry-adapted operators that incorporate problem symmetries such as the conservation of spin, particle number, and the point-group.
Ref.~\cite{OperatorTiling2024} introduces operator tiling for translation-symmetric systems, tiling problem-relevant Pauli operators chosen by ADAPT-VQE across the entire system.
Ref.~\cite{ramoa2024reducingresourcesrequiredadaptvqe} introduces Coupled Exchange Operators which create linear combinations of qubit-excitation operators acting on the same set of spin-orbitals.

Of particular interest to high energy and nuclear physics are the Scalable-Circuits \Adapt (\SCAdapt) for extrapolating ansätze from small to large systems~\cite{Farrell:2023fgd, Farrell:2024fit}. This method adopts a pool of coordinate-invariant operators, so that circuits trained on a series of classical \Adapt simulations at small volumes can be scaled to those at large volumes, which 
are then run on a quantum computer. This obviates the need for running \Adapt or even VQE to obtain optimized parameters on a large volume, since the parameters are obtained via fitting and extrapolation from smaller systems. 
The operator pool for \Adapt is arbitrary, in principle,
but 
judicious choices for the pool operators can improve convergence. 
The operator pool used in Ref.\,\cite{Farrell:2023fgd} for running \Adapt on the Schwinger model Hamiltonian in \Eq{schwingerham}, which we will refer to as the {\em full singles} pool, 
 includes 
\begin{subequations} \label{eq:O:pool}
    \begin{align}
    \hat{O}^V_{mh}(d)  = & \sum_{n=0}^{N_s - 1 - d} (-1)^n 
    \hat{O}_{n}^{(d)},   \label{eq:Omh:V}
 \\
    \hat{O}^S_{mh}(m, d)& =  
    \bigl(
       \hat{O}_{m}^{(d)}  + \hat{O}_{N_s-m-1-d}^{(d)} 
    \bigr),
    \label{eq:Omh:S} 
    \end{align}
where
\begin{align}
    \label{eq:source}
    \hat{O}^{(d)}_{n} \equiv \frac{i}{2}\Big(\hat{\sigma}^+_n \hat{Z}_n^d \hat{\sigma}^-_{n + d} - \rm{h.c.}\Big) \; ;
\end{align}
\end{subequations}
$\hat{Z}_n^d \equiv \bigotimes_{q} \hat{Z_q}$
denotes a string operator of length $d-2$ acting on all sites $q$ between
$\sigma_n^+$ and $\sigma_{n+d}^-$.
The subscript with $\hat{O}_{mh}$ stands
for the combined effect of $m$ass and $h$opping terms
\cite{Farrell:2023fgd}.
The superscripts $V$ or $S$ indicate that the operator is a volume (bulk) or
surface (boundary) term, respectively. 
These terms respect the time reversal 
symmetry of the ground state of the model, and
represent 
the insertion of a single meson of length $d$.
While \Eq{Omh:V} 
does so uniformly throughout the system,
\Eq{Omh:S} only does so at
a distance $m$ from the open boundaries.
The latter
is expected to be useful, as boundary effects will be critical regardless of the values of 
$g$ and $m_0$. Of importance here is that the operator pool construction leads to a different set of operators being available at each volume. When the \Adapt trajectory includes all new pool operators appearing at each volume, robustly extrapolating the sequence of operators becomes non-trivial.

The insertion of single meson is like  
the singles term in a unitary coupled cluster (UCC) ansatz. 
When including all permissible
$d$ and $m$ for a given system size $N_s$ in \Eqs{O:pool},
henceforth we refer to this as the \textit{full singles pool}.
This method can be extended to include an ansatz which
inserts two mesons which do not overlap, or double terms
\cite{peruzzo2014variational,lee2018generalized} which are of the form $i\hat{a}^\dagger_j \hat{a}_{j+d}\hat{a}^{\dagger}_k\hat{a}_{k+d'} - h.c.$.
For the parameters we studied, however, we found that the contribution
of such two meson operators was negligible. 
It is not clear if this is true in general, especially for models that possess many internal degrees of freedom.
One notable example is that of multiple flavors of quarks in quantum chromodynamics. 

\section{\newAdaptName}
\label{sec:sc2}

Our algorithm, which we term Surrogate-Constructed Scalable-Circuits \Adapt, or \newAdaptName, uses a classically tractable {\em surrogate} theory to select a subset of relevant operators in the \Adapt operator pool, independent of volume.
This procedure defines a new, {\em truncated} pool, and a {\em minimum volume} for which {\em any} ADAPT-VQE circuit built from the truncated pool is well-defined.
Therefore, we construct an ansatz just once, by running a classical \Adapt simulation with the truncated pool on the largest tractable volume.
To generate a series of circuits from which to extrapolate to an even larger volume simulable with a quantum computer,
    we run classical VQE simulations {\em with a consistent ansatz} on each smaller volume, down to the minimum volume defined by the truncated pool.
The high level flow chart for this algorithm is illustrated in Fig. \ref{fig:daflowchart}.

\begin{figure}
\includegraphics[width=\linewidth]{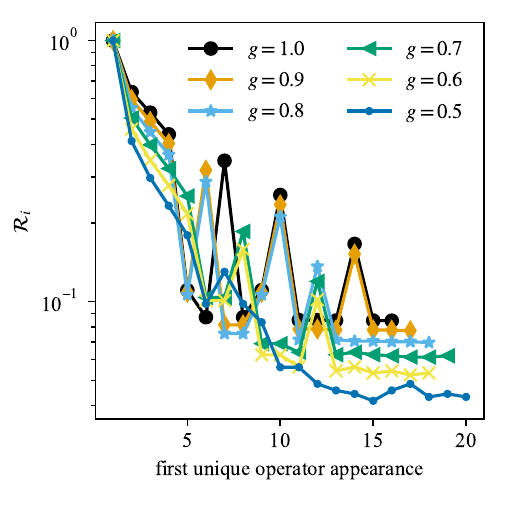}
\caption{Scaling of the transition matrix element operator, $\mathcal{R}_i$, versus the order the operator first uniquely appears in ADAPT-VQE applied to the massless Schwinger model, 
using the pool in Eq.~\ref{eq:O:pool} on lattices with $N_s=16$ and various couplings $g$ as specified in the legend. 
Operators that appear later suggest that they are less relevant to the state preparation. We see that it indeed does strongly correlate with $\mathcal{R}_i$. }
\label{fig:operatortrends}
\end{figure}

\begin{figure*}
\includegraphics[width=\linewidth]{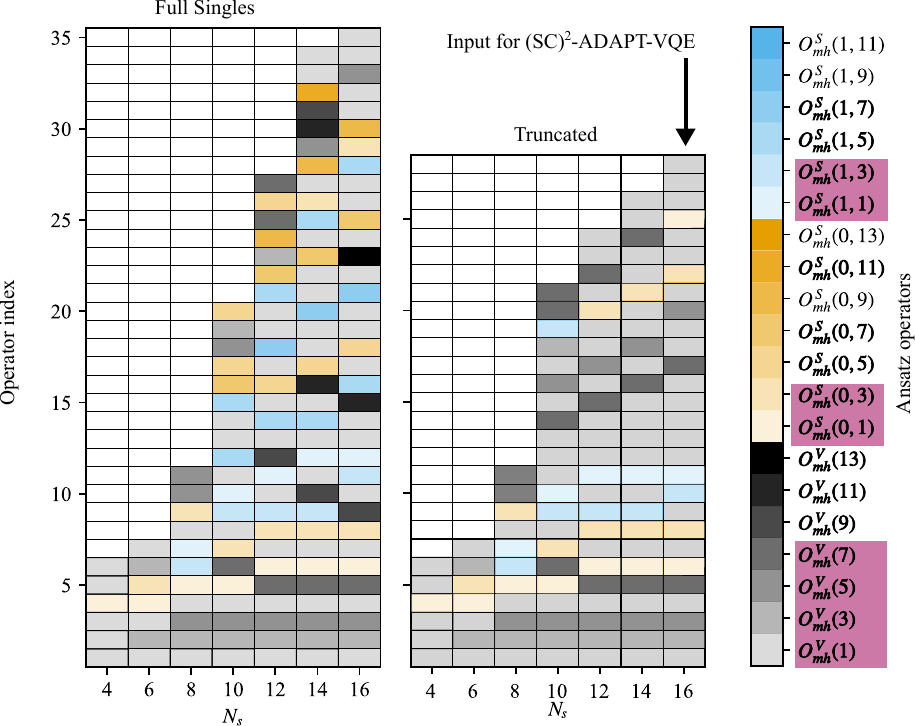}
    \caption{Ansatz order for the full singles pool (left) and the truncated pool (right), using \Adapt with  $g=1.0$
    and $m_{0}=0.0$. These simulations used the gradient convergence criterion of $\epsilon = 10^{-3}$, \Eq{adapteps}.
    The operators are defined in Eq. (\ref{eq:O:pool}),
    while the legend to the right is the full singles pool for $N_s =16$.  
    Bold-face in the legend indicates that the operator was selected by \Adapt using the full singles pool for at least one volume.
    The magenta highlighting indicates which operators were chosen for the truncated pool with $\Delta=10^{-5}$.
    The different colors denote different types of operators that appear in the ansatz.  Volume operators are shaded from
    grey to black as their length increases.  Orange denotes surface operators that touch the edge of the lattice, while blue operators are one lattice site from the edge. Notice that there is a striking regularity of the operators which persist as $N_s$ increases for                  
    the surrogate truncated pool, versus that of the full singles pool.  This demonstrates the utility of our truncation method.
    }
    \label{fig:ansatzstabilitytrunc}
\end{figure*}

The surrogate theory can widely vary, but it should be accurate, reasonably scalable, systematically improvable, and 
calculable efficiently at the classical level. Examples of such surrogate theories could include coupled cluster calculations, matrix product states, sparse wavefunction simulators, and Markov chain Monte Carlo.
The first step of \newAdaptName uses a surrogate theory to measure the overlap operator
\begin{align}
    \label{eq:Mj}
    \mathcal{M}_j \equiv |\langle \psi_{\rm{ref}}| \hat{O}_j
    \grdapx 
    |^2.
\end{align}
In this equation $\hat{O}_j$ is the Hermitian operator that generates the rotations in Eq. (\ref{eq:vqeansatz}), and 
$\grdapx$ 
is an approximation of the ground state that might be determined classically.
This can be done in several ways:
in $1+1$ dimensions, using the density matrix renormalization group,
or for higher dimensions, with projected pair entangled states. Lastly, $|\psi_{\rm{ref}}\rangle$ is an initial state that is easy to prepare. Here we use the strong coupling limit for the Schwinger model ($g\rightarrow\infty$), $|\psi_{\rm{ref}}\rangle=|0101.....01\rangle$. In chemistry problems one might use the Hartree-Fock state for
$|\psi_{\rm{ref}}\rangle$.
The values $\mathcal{M}_j$ are calculated only once for a choice of couplings $(g$, $m_0)$ before any \Adapt calculations are run. In particular it is worth noting that the states themselves
$\grdapx$ 
and $|\psi_{\rm{ref}}\rangle$ do not need to be calculated explicitly, only the transition matrix element $\mathcal{M}_j$.

This operator is similar to a transition matrix element as it is the square of the linear term in the Maclaurin series approximation of 
$\langle \psi_{\rm ref}|e^{-i\theta \hat{O}_j} \grdapx$, up to an overall constant.
This choice of operator is reasonably well founded as it forms an approximation for the leading order contribution to a Maclaurin series expansion of the target operator. This transition matrix element is calculated classically. 
Next we impose a cut-off $\Delta$ such that we only keep operators $\hat{O}_j$ that satisfy 
\begin{equation}
    \mathcal{R}_j \equiv \frac{|\mathcal{M}_j|}{\rm{max}(|\mathcal{M}_j|)} \geq \Delta.
    \label{eq:Rratio}
\end{equation}
While the fixed choice of the cut-off, $\Delta$, is 
arbitrary,
it allows for a consistent algorithmic construction of the operator pool, by removing possible biases at a given choice in parameter space. As $\Delta$ is taken to smaller values the size of the operator pool increases. 

The operators chosen in standard ADAPT-VQE, and the scaling of 
$\mathcal{R}_j$, are correlated,
as illustrated in Fig. \ref{fig:operatortrends}. 
We find that the MPS amplitudes behave in a reasonably correlated manner with respect
to the order of their first appearance in the ADAPT-VQE ansatz.

\begin{figure}[!ht]
    
    \includegraphics[width=\linewidth]{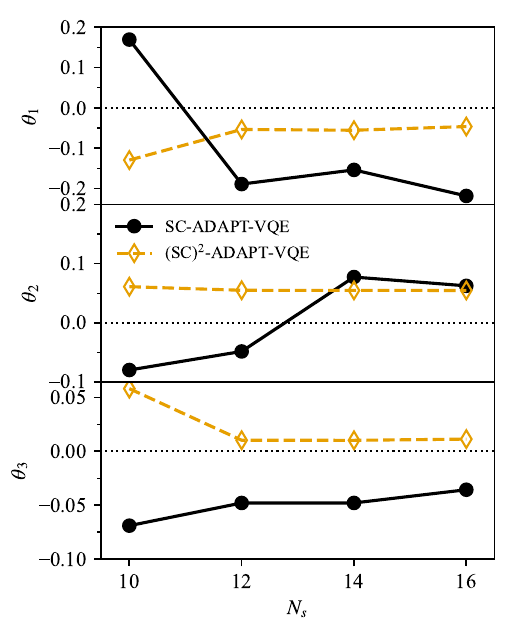}
    \vspace{-3em}
    \caption{
    Optimized coefficients $\theta_j$ for the first (top), second (middle), and third (operator) selected by \Adapt applied to the massless Schwinger model with $ag=1.0$. For the \SCAdapt curve (solid), \Adapt was solved at each volume using the pool in Eq.~\ref{eq:O:pool}, up to a gradient convergence criterion of $\epsilon=10^{-3}$. For the \newAdaptName curve (dashed), \Adapt was solved at a volume of $N_s=16$ up to the same $\epsilon$ and using the same pool, but truncated according to Eq.~\ref{eq:Rratio} with $\Delta=10^{-5}$. The resulting circuit was re-optimized for volumes $N_s<16$.
    All runs of \Adapt selected the same first three operators, irrespective of volume ($O_{mh}^V(1)$ for $\theta_1$, $O_{mh}^V(3)$ for $\theta_2$, $O_{mh}^V(5)$ for $\theta_3$). However, the \SCAdapt runs at different volumes selected different operators later on in the ansatz.
    This produced an extreme non-monotonicity in the optimized coefficients, and so an ambiguous extrapolation to the 
    limit of infinite volume. The \newAdaptName protocol resolves this problem.
    }
    \label{fig:opvalset}
\end{figure}

One source of instability in \SCAdapt is the relatively large volume needed to obtain a consistent operator sequence in each run of \Adapt. 
This is illustrated for $g=1.0$ and $m_0= 0$ in
\Fig{ansatzstabilitytrunc} for both the truncated ($\Delta > 10^{-5}$) and full singles ($\Delta=0$) operator pool. 
The figure uses shaded boxes to indicate different operators in the circuit: the grey operators correspond to the volume operators, $\hat{O}^V_{mh}(d)$, while the blue and orange regions correspond to the surface or boundary operators, $\hat{O}^S_{mh}(m, d)$.
While operators with an early depth, {\it i.e.} the first through eighth,   stabilize by $N_s=12$, at this point the later operators have not stabilized. This implies that for progressively deeper (more accurate) ans\"atze, one needs progressively larger classical calculations to find a quantum
circuit with an appropriately large volume. 
Crucially, we observe that the presence of different operators later on, which are needed for ADAPT to converge, still results in erratic behavior for the parameters ofof those early operators.  
Using the ansatz for $10\leq N_s \leq 16$, we show mild non-monotonicity for the optimization angles for the first three operators in Fig. \ref{fig:opvalset}. 
With \newAdaptName, we immediately start with the largest volume
at which we can simulate classically, which we fix at $N_s=16$.
We use this ansatz to set the operator ordering, and then, where
feasible, use the same circuit structure at smaller volumes.
Doing so, we find that the optimized parameters exhibit increased
monotonicity. One feature of note is that given the various operators $\hat{O}_{mh}^V(d)$ are themselves Trotterized, it is possible for \Adapt to select the same operator multiple times which might not typically happen in non-lattice systems. 

An obvious limitation of \newAdaptName is the capabilities of the surrogate theory. This has to be kept in mind 
particularly for critical systems, or when taking the continuum limit, as then the correlation length of the system,
in units of the lattice spacing, diverges.  This places limits on the range of parameters directly accessible from a given surrogate. Alternatively, it places constraints on the type of surrogates 
that should be used.

\section{Results}
\label{sec:results}
While constructing a gauge theory on a lattice is a useful computational tool, 
eventually we are interested in the continuum limit.
Since the product of the lattice spacing times the gauge coupling, $ a g$,
is dimensionless, at fixed lattice spacing $a$ we can approach the
continuum limit by taking $g \rightarrow 0$.  
In doing so, the ratio of $m_0/g$ is held fixed, with
different values corresponding to distinct theories.

In this section we show how results using \newAdaptName can
be used to reliably extrapolate to the continuum limit
for the massless theory.  While usually this is the most challenging case,
for the special case of the Schwinger model the model is solvable
analytically, with the spectrum only  
a single free, massive boson.
As a fiducial quantity, we compute the expectation value of the chiral condensate,
\begin{align}
    \langle \bar{\psi}\psi\rangle = \frac{1}{2 N_s} \sum_{j=0}^{N_s-1} (-1)^j\langle\hat{\sigma}^z_j\rangle. 
\end{align}
This expectation value is trivially nonzero
when there is a fermion mass.
However, it remains non-zero even if $m_0 \to 0$.
There, 
spontaneous dimerization of the system occurs dynamically,
with the analytical continuum limit given by\cite{Nielsen:1976hs}
\begin{eqnarray}
    \langle \bar{\psi}\psi\rangle _{0} 
    &=& -\frac{g}{2\pi^{3/2}}e^{\gamma_E}
    \qquad (m_0=0)
\label{eq:continuumSchwinger}
\end{eqnarray}
with $\gamma_E=0.5772$  
the Euler constant.

We choose to compute $\langle \bar{\psi}\psi\rangle$ rather than just
the energy (which serves as the objective function in our \Adapt calculations), as it receives contributions from
off-diagonal elements in the energy eigenbasis, and so provides a better
measure of accuracy. 

\begin{figure}
\includegraphics[width=\linewidth]{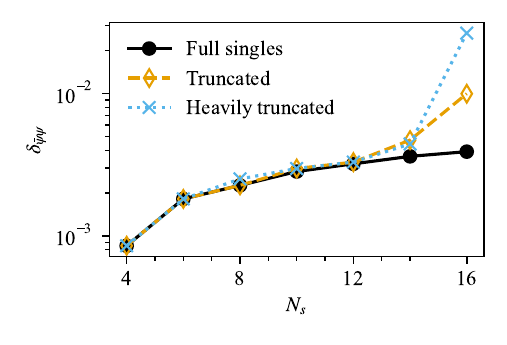}
\caption{
Relative error 
$\delta_{\bar\psi\psi} \equiv
\qty|\expval{\bar\psi\psi}-\expval{\bar\psi\psi}_{0}|/\qty|\expval{\bar\psi\psi}_{0}|$ 
of chiral condensates 
obtained from \newAdaptName applied to the massless lattice Schwinger model with $ag=0.7$
using \Eq{continuumSchwinger}.
\Adapt was solved at each volume up to a gradient convergence criterion of $\epsilon=10^{-3}$. Each curve used a distinct pool, constructed from Eqs.~(\ref{eq:O:pool}) and~(\ref{eq:Rratio}) with $\Delta=0$ (full singles), $10^{-5}$ (truncated), and $10^{-3}$ (heavily truncated).
Chiral condensates for each optimized circuit were measured, and their relative error was calculated with respect to DMRG calculations of matching volume, with a maximal bond dimension of 400.
}
\label{fig:chiralcompcomparison}
\end{figure}

Before proceeding to the continuum limit,
it is 
important to investigate aspects of the operator pool truncation at finite volumes and non-zero lattice spacings. 
Fig. \ref{fig:chiralcompcomparison} shows the relative error of the chiral condensate at $m_0=0$ and $g=0.7$, of distinct \Adapt trial states constructed independently for each volume, using
three fiducial values of the threshold $\Delta$ for $\mathcal{R}_j$:  $\Delta=0,~10^{-5},$ and $10^{-3}$, which we denote as the
full singles, truncated, and heavily truncated operator pools, respectively. 
For sufficiently small volumes there is no significant difference between the three truncations. 
By a volume of $N_s=14$, though,
significant discrepancies arise, and it is clear that there are ``truncation" artifacts that become noticeably important. 
\begin{table}[t]
\caption{
Characteristic numbers for \newAdaptName applied to the massless Schwinger model, using pools constructed from Eqs.~\ref{eq:O:pool} and~\ref{eq:Rratio} with different values of $\Delta$. 
The ``Depth'' columns identify the number of operators selected by \Adapt at a volume of $N_s=16$ up to a gradient convergence criterion of $\epsilon=10^{-3}$.
The ``Min $N_s$'' columns identify the minimum volumes for which all operators in the pool are defined.
Without truncation (i.e. using $\Delta=0$), there is no well-defined minimum $N_s$.
}
\label{tab:params}
\begin{tabular}{c|c|c|c|c|c}
\hline\hline
& \multicolumn{1}{c|}{\makecell{Full singles \\ ($\Delta=0$)}}
& \multicolumn{2}{c|}{\makecell{Truncated \\ ($\Delta=10^{-5}$)}}
& \multicolumn{2}{c}{\makecell{Heavily truncated \\ ($\Delta=10^{-3}$)}} \\
\hline
$ag$ & Depth & Depth & Min $N_s$ & Depth & Min $N_s$ \\
\hline\hline
1.0 & 34 & 27 & 10 & 22 & 6 \\
0.9 & 32 & 27 & 10 & 23 & 6 \\
0.8 & 35 & 27 & 10 & 22 & 6 \\
0.7 & 31 & 32 & 12 & 32 & 8 \\
0.6 & 33 & 36 & 12 & 30 & 8 \\
0.5 & 37 & 36 & 14 & 32 & 10 \\
\hline\hline
\end{tabular}
\end{table}
\begin{figure}
\includegraphics[width=\linewidth]{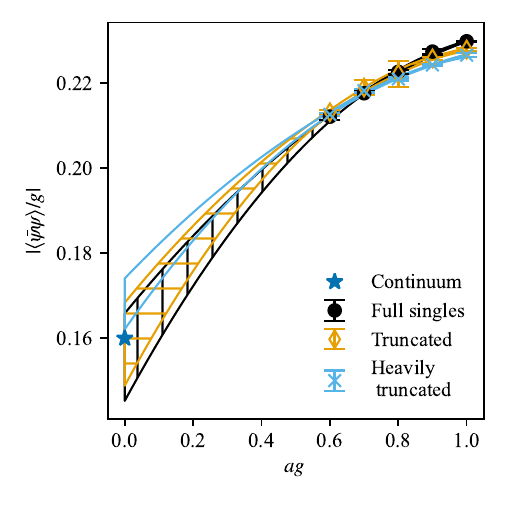}
\caption{
Chiral condensates obtained from \newAdaptName applied to the massless lattice Schwinger model, with an extrapolation to the continuum limit. Each curve used a distinct pool, constructed from Eqs.~\ref{eq:O:pool} and~\ref{eq:Rratio} with $\Delta=0$ (full singles), $10^{-5}$ (truncated), and $10^{-3}$ (heavily truncated). In the truncated and heavily truncated curves, \Adapt was solved for each parameter value $ag$ at a volume of $N_s=16$ up to a gradient convergence criterion of $\epsilon=10^{-3}$, and the resulting circuits were re-optimized for volumes $N_s<16$, down to the minimum volumes given in Table~\ref{tab:params}. Because there is no defined minimum volume for the full singles pool, \Adapt was solved independently for each volume $4\le N_s\le16$. Chiral condensates for each optimized circuit were measured, and extrapolated to the thermodynamic ($N_s\rightarrow\infty$) limit, yielding the points seen in the plot.
 The errorbars with the symbols derive from
 systematic errors occurring from the infinite volume extrapolations.
The shaded areas 
show an extrapolation to the continuum limit ($a\rightarrow0)$
as the confidence interval of the underlying polynomial fit.
The asterisk at $ag=0$ represents the exact analytic result, 
\Eq{continuumSchwinger} for $g=1$.
This demonstrates that the 
truncation of the operator pool has minimal effect
on the final continuum limits.}
\label{fig:continuumlimitchiralcondensate}
\end{figure}
The deviations grow weakly towards
larger volumes where certain operators transition from being irrelevant to important. These cases will become more prevalent when the gradient truncation threshold $\epsilon$ is decreased, and volume artifacts begin to stabilize.

We are now in a position to investigate how truncating the operator pool affects calculations in the 
thermodynamic and continuum limits.
In Fig. \ref{fig:continuumlimitchiralcondensate}, we perform the \newAdaptName procedure (solving \Adapt at just one volume for each coupling parameter) for the truncated and heavily truncated pools to obtain estimates of the chiral condensate in the thermodynamic limit. As a control, we compare against the \SCAdapt procedure (solving \Adapt for each volume) with the full singles pool. Note that for these results, we extrapolate the observable directly, rather than the operator coefficients, to sidestep the instabilities observed in Fig.~\ref{fig:opvalset}.

Using the thermodynamic limit values for each coupling parameter, we then perform another extrapolation to the continuum limit.
For $g=1$, 
from \Eq{continuumSchwinger} the result 
in the continuum limit is $\chiconexp_{0}=- 0.159929$. 
Using \newAdaptName with a truncated pool, the
result we find after taking $a g \to 0$,
 indicated by the
blue asterisk in \Fig{continuumlimitchiralcondensate},
agrees well with this result.
This demonstrates that while certain marginal operators may appear in the full \Adapt pool, that at least for
the value of the chiral condensate, in practice their
effect is minimal.
This robustness encourages using \newAdaptName
to reach the continuum limit in this and other field theories.

\section{Outlook}
\label{sec:outlook}
We have developed an algorithm, \newAdaptName, which 
addresses certain ambiguities in the
\SCAdapt
algorithm of Ref. \cite{Farrell:2023fgd}.
This method utilizes classically scalable resources to efficiently identify an operator pool
which is well-defined on any number of qubits above a minimum volume.
This allows a single classical execution of ADAPT-VQE to generate a fixed ansatz which can be optimized via VQE for smaller volumes, providing a robust sequence of parameters that can be extrapolated to large volumes executable on quantum hardware.
We also note the reduced operator pool can reduce the overall sampling budget by
a factor linear in the size of the system.
We have not examined our extensions to the algorithm at scale on a quantum computer because the noise on present day devices prohibitively accumulates for the long circuit depths we consider.
Nevertheless, this improvement to the algorithm 
greatly stabilizes extrapolating to larger, scalable circuits.
This is crucial for looking at theories with long correlation lengths,
especially as arises in approaching the continuum limit.
In particular when we want to expand to larger systems, the choice of $\Delta$ can depend on the process, observable, and model or theory. Therefore, it will be important to develop other theory-specific heuristics that might improve upon or replace the choice procedure of the cutoff. 

We have demonstrated that \newAdaptName can be used to approach the continuum limit of relativistic field theories, using the Schwinger model as an example.  This makes it a promising method for state preparation of near-term quantum simulations of a variety of low-dimensional models, including the multi-flavor Schwinger model \cite{Coleman:1976uz}, QCD$_2$ \cite{tHooft:1974pnl}, and even bosonic models such as $\lambda \phi^4$ \cite{Glimm:1968kh} and the Abelian-Higgs model \cite{Rubakov:1985ehm}.  A particularly interesting prospect would be useful to apply this method to
QZD, a $Z(3)$ gauge theory coupled to three flavors of fermions
\cite{Florio:2023kel}.  This model has both confined
fermions and bosons, analogous to the baryons and mesons of QCD.
Like QCD, cold, dense QZD has a severe sign problem. Using
\newAdaptName to study QZD will be of use both to investigate the Fermi sea of the theory, as well as to understand how \newAdaptName behaves in the presence of more complicated excitations. 

The increase in correlation length of the system as the continuum is approached is only a specific example of a system going through a second-order phase transition. As such, \newAdaptName also poses itself as an ideal algorithm to address the issue of ground state preparation of critical systems, close to and away from quantum phase transitions. This includes many models in condensed matter physics, such as the Hubbard model.

In \Adapt simulations of molecular systems utilizing Gaussian basis sets, one could easily utilize portions of the \newAdaptName algorithm for the operator pool selection component. In particular this method should be able to reduce 
the number of possible operators for an
UCC type ansatz, which is useful in electronic systems. 

\begin{acknowledgements}

The authors thank Anthony Ciavarelli, Roland Farrell,  Marc Illa, Abid Kahn, Wayne J. Mullinax,  Martin Savage, and Nikita Zemlevskiy for helpful comments.  K. Sherbert, A.F., K. Shirali, 
Y.C., S.V., A. W., S.E.E., R.D.P., and N.M.T. are supported by the U.S. Department of Energy, Office of
Science, National Quantum Information Science Research
Centers, Co-design Center for Quantum Advantage (C$^2$QA)
under Contract No. DE-SC0012704.  C$^2$QA led
this research.  E.G. was supported by the NASA Academic Mission Services, Contract No. NNA16BD14C and NASA-DOE interagency agreement SAA2-403602.  A.F. and R.D.P. were supported by the U.S. Department of Energy under contract DE-SC0012704. H.L. and E.G., are supported by the U.S. Department of Energy, Office of Science, National Quantum Information Science Research Centers, Superconducting Quantum Materials and Systems Center (SQMS) under contract number DE-AC02-07CH11359. Fermilab is operated by Fermi Research Alliance, LLC under contract number DE-AC02-07CH11359 with the United States Department of Energy. S.V. was supported by
the U.S. Department of Energy, Nuclear Physics Quantum Horizons program through the Early Career Award DE-SC0021892.
A.W. was supported by the U.S. Department of Energy, Office of Science, Basic Energy Sciences, Materials Sciences and Engineering Division. 
 The authors acknowledge use of the HSL4 computing system for larger scale simulations.
  This research used resources of the National Energy Research
Scientific Computing Center, a DOE Office of Science User Facility supported by the Office of Science of the U.S. Department of Energy using NERSC award ASCR-ERCAP0024469.

\end{acknowledgements}

%

\end{document}